\documentclass[12pt]{article}
\usepackage{epsf}

\newcommand{\eeq}{\end{equation}}
\newcommand{\vs}[1]{\rule[- #1 mm]{0mm}{#1 mm}}
\newcommand{\beq}{\begin{equation}}
\newcommand{\beql}{\begin{eqnarray}}
\newcommand{\eeql}{\end{eqnarray}}

\newcommand{\lam}{\lambda}

\newcommand{\mP}{\mathcal{P}}
\newcommand{\mS}{\mathcal{S}}
\newcommand{\mR}{\mathcal{R}}
\newcommand{\mW}{\mathcal{W}}
\newcommand{\mY}{\mathcal{Y}}
\newcommand{\mV}{\mathcal{V}}

\newcommand{\sect}[1]{\setcounter{equation}{0}\section{#1}}
\renewcommand{\theequation}{\thesection.\arabic{equation}}

\begin{document}

\begin{titlepage}

\vs{5}

\begin{center}

{\LARGE {\bf Chiral Random Matrix Models:\\[.5cm] 
A Novel Intermediate Asymptotic Regime}}\\[2cm]
             
{\Large N. Deo$^{1,2}$}\\
{$^1$ Poornaprajna Institute of Scientific Research}\\
{4 Sadashiva Nagar, Bangalore 560080, India}\\
{$^2$ Abdus Salam International Center}\\
{for Theoretical Physics, Trieste, Italy}\\
{ndeo@vsnl.net}\\

\end{center}

\vs{10}

\centerline{ {\bf Abstract}}

\noindent

The Chiral Random Matrix Model or 
the Gaussian Penner Model (generalized Laguerre ensemble)
is re-examined in the light of the results
which have been found in double well matrix models \cite{D97,BD99} 
and subtleties discovered in the single well matrix models \cite{BH99}.
The orthogonal polynomial method is used to extend the 
universality to include non-polynomial potentials. 
The new asymptotic ansatz is derived (different from Szego's
result) using saddle point techniques. 
The density-density correlators are the same
as that found for the double well models ref. \cite{BD99}
(there the results have been derived for arbitrary potentials). 
In the smoothed large $N$ limit they  
are sensitive to odd and even $N$ where $N$ is the size
of the matrix \cite{BD99}. This is a more realistic 
random matrix model of mesoscopic systems with density 
of eigenvalues with gaps. The eigenvalues see a brick-wall 
potential at the origin.
This would correspond to sharp edges in a real mesoscopic
system or a reflecting boundary. Hence the results
for the two-point density-density correlation function may
be useful in finding one eigenvalue effects in 
experiments in mesoscopic systems or small metallic grains. 
These results may also be relevant for studies of structural
glasses as described in ref. \cite{D02}.


\end{titlepage}

\renewcommand{\thefootnote}{\arabic{footnote}}
\setcounter{footnote}{0}

\sect{Introduction}\label{intro}

In this work we re-visit the Penner model in the context of disordered 
mesoscopic systems where they naturally appear 
e.g. in disordered models of metal-insulator transitions \cite{CIM92,SN94},
superconducting-normal interface with Andreev reflection 
(the openning of a gap) \cite{B97}
and also as models of structural glasses \cite{D02}.  
Historically the Penner matrix model was studied in the context
of the moduli space of a punctured surface \cite{PHZ86}. There  
an equality between the Penner matrix model and the Euler characteristic
of moduli space of punctured surfaces was computed, before taking 
the continuum limit. Later in \cite{DV91} it was shown that even after
taking the continuum limit the Euler characteristic of moduli
space of unpunctured surfaces was obtain as the free energy of
the Penner model. This was done in an effort to understand  
2-dimensional gravity coupled to matter at the critical point 
c=1. Recent advances in this direction are contained in \cite{DV02}.
There has also been recent work on the generalized Laguerre
ensemble in the context of the Chiral Random 
Matrix models of QCD (see \cite{A02,FS02} and references therein)
and in describing a novel group structure associated
with scattering in disordered mesoscopic wires \cite{SN94}. 

This study focuses on the Gaussian Penner model 
with potential $V(M)={1\over 2} \mu M^2 - {t\over 2} \ln M^2$
where M is a $N\times N$ random matrix; after some work on gapped 
matrix models has been reported and clarified \cite{D97,BD99,D02}. 
In the recent work two basic ideas have been implemented. 
First the idea of symmetry breaking (see \cite{BDJT93,D02})
and then a new asymptotic ansatz of the orthogonal
polynomials \cite{D97}. Here for the Gaussian Penner model the corresponding 
polynomials are the associated Laguerre polynomials, $L^{\alpha}_N (x)$,
where both the $\alpha$ and $N$ asymptotes are to be taken,
in previous work only the $N$ asymptote was found. It
turns out that it is in this asymtotic region that the singular 
models make contact with the double well random matrix models and 
the universality of the gapped matrix model is extended to include
non-polynomial potentials. Ideas of symmetry breaking are also true 
for the Gaussian Penner model \cite{BDJT93,D02} but because of the brick 
wall potential at the origin, which involves one electron quantum 
tunneling through an infinite barrier, it is harder to give an 
intuitive picture. These ideas will be ellaborated on in future papers.

The paper is divided as follows. It starts by
describing completely the model and establishing
the notation and conventions. In section 2, the old asymptotic
ansatz of Szego for the associative Laguerre polynomials
is taken and shown to not correspond to the new
universality of the double well matrix model studied in \cite{D97,BD99}.
In section 3, the new asymptotic ansatz is derived using saddle point
techniques. This corresponds to the asymptotic ansatz discussed in
\cite{D97,BD99}. Section 4 ends with conclusions and open
questions. 

\sect{The Model, Notation and Conventions}\label{notation}

\begin{figure}     
\leavevmode
\epsfxsize=4in
\epsffile{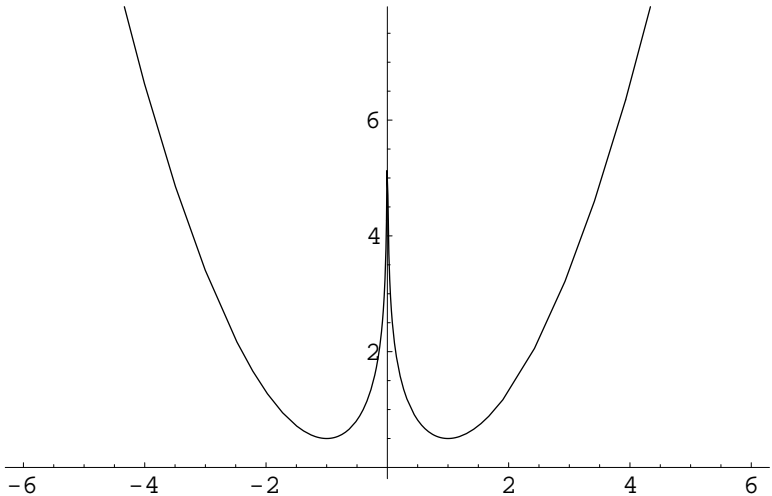}
\caption{The potential for the Gaussian Penner Random Matrix Model
with $V(M)={1 \over 2} \mu M^2 - {t\over 2} \ln M^2$,
$\mu=1$ and $t=1$.}
\label{penner} 
\end{figure}

We consider models of the type (see \cite{T92} for details of
notation and definitions)

\beq
Z = exp{(-F)} = \int dM e^{-N Tr V(M)}
\eeq

where $V(M)=V_0 (M) - {t\over 2} ~~ \ln M^2$ and here $V_0 (M)
= {1\over 2} \mu M^2$. See Fig. \ref{penner}.

In general the orthogonal polynomials are $P_n (\lam) = \lam^n + l.o.$
where $\lam$ are the eigenvalues of $M$ and the orthogonality conditions 
are $\int_{-\infty}^{\infty}d\lam e^{-NV(\lam)}P_n(\lam)P_m(\lam)
=h_n \delta_{nm}$. The partition function can be expressed in terms
of the $h_n$'s as $Z=N!h_0h_1h_2......h_{N-1}$.

\begin{figure}     
\leavevmode
\epsfxsize=4in
\epsffile{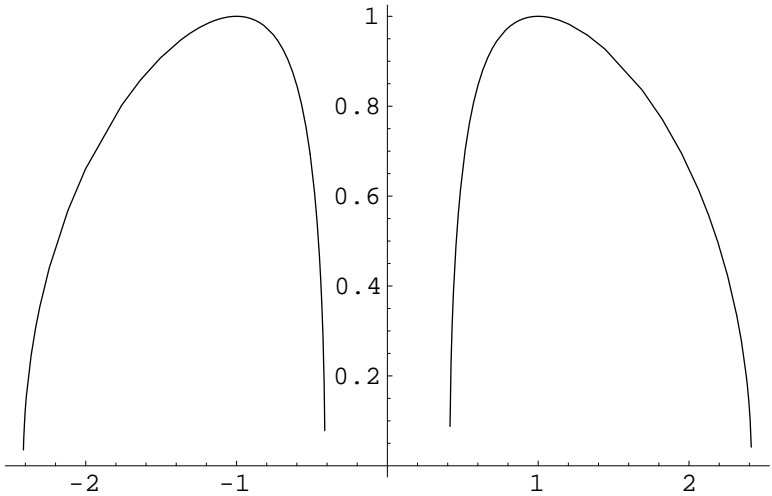}
\caption{The density of eigenvalues for the 
Gaussian Penner Random Matrix Model for $\mu=1$ and $t=1$.}
\label{pennerdos} 
\end{figure}

For the large N limit, the density of eigenvalues, $\rho(z)\equiv
({1\over N}) \sum_{i=1}^N \delta(z-\lambda_i)$, can be found by 
solving either a saddle point equation or the Schwinger-Dyson equation.  
In terms of the generating function $F(z)={1\over N}
\left[ tr{1\over z-M} \right] \rightarrow \int {dz^{\prime}}
{\rho(z^{\prime})\over {z-z^{\prime}}}$ the Schwinger-Dyson
equation reads $F(z)^2 - V^{\prime} (z) F(z) = M(z)$, where $M(z)$
is a meromorphic function. The density of eigenvalues is given by
$\rho(z)=-{1\over \pi} Im F(z)$. The
generating function $F(z)$ in the large N limit for $t>0$ is
\beq
F(z) ={ {\mu z}\over 2} - {t\over {2 z}} - 
{\mu \over {2 z}}\sqrt{(z^2-a^2)(z^2-b^2)} 
\eeq
where $a^2={(2+t)\over \mu}+{2\over \mu}\sqrt{(1+t)}$ 
and $b^2={(2+t)\over \mu}-{2\over \mu}\sqrt{(1+t)}$.
See Fig. \ref{pennerdos} for the corresponding density of eigenvalues.

\sect{Exact Solution for the Symmetric Gaussian Model}\label{exact}
 
The orthogonal polynomials satisfy the recurrence relation 

\beq
(z-S_n) P_n (z) = P_{n+1} (z) + R_n P_{n-1} (z)
\label{rec}
\eeq
where $S_n, R_n$ are the recurrence coefficients.
For symmetric models $S_n=0$. For even potentials instead of
recurrence relation 

\beq
z P_n (z) = P_{n+1} (z) + R_n P_{n-1} (z)
\label{pzero}
\eeq

one can use

\beq
z^2 P_{2n} (z) = P_{2n+2} (z) + (R_{2n+1}+R_{2n}) P_{2n} (z)
+ R_{2n-1} R_{2n} P_{2n-2} (z)
\label{pone}
\eeq

and 

\beq
z^2 P_{2n+1} (z) = P_{2n+3} (z) + (R_{2n+1}+R_{2n+2}) P_{2n+1} (z)
+ R_{2n+1} R_{2n} P_{2n-1} (z).
\label{ptwo}
\eeq

Where we have multiplied z to Eq. (\ref{pzero}) and expanded. Then
eq. (\ref{pone}) contains only even polynomials
$P_{2n} (-z) =  P_{2n} (z)$ and 
eq. (\ref{ptwo}) contains only odd polynomials
$P_{2n+1} (-z) = - P_{2n+1} (z)$. This simplifies the solution
as we will see.

(1). Let us first work with the even set. Let $y=z^2$ and define
functions $\mathcal{P}_{n} (y) = P_{2n} (z)$. In terms of these 
`new' polynomials

\beq
y \mP_n (y) =  \mP_{n+1} (y) + \mS_n \mP_n (y) + \mR_n  \mP_{n-1} (y)
\eeq

where $\mS_n = R_{2n} + R_{2n+1} = A_n + B_n$ and
$\mR_n = R_{2n-1} R_{2n} = A_n B_{n-1}$. These polynomials obey

\beq
\int_0^{\infty} dy e^{-N^{\prime} [ \mV_0 (y) - t^{\prime} log y ]}
\mP_n (y) \mP_m (y) = h_n \delta_{n,m}
\eeq
where
$
\mV_0 (y) = 2 V_0 (z) = \mu y + .......
$, $t^{\prime} = t - {1\over {2 N^{\prime}}}$ and
$N^{\prime} = {N\over 2}$.

This is the same as the brick wall problem i.e. the linear
Penner model if $t \leftrightarrow t^{\prime}$, 
$N \leftrightarrow N^{\prime}$ and $t^{\prime} N^{\prime} 
= (t-{1\over N}) {N\over 2} = {(Nt-1)\over 2}.$

(2). A similar analysis can be carried out for the odd set. Define

\beq
\bar{\mP}_n (y) = z^{-1} P_{2n+1} (z).
\eeq

Then

\beq
y \bar{\mP}_n (y) = \bar{\mP}_{n+1} (y) + \bar{\mS}_n \bar{\mP}_n (y)
+ \bar{\mR}_n \bar{\mP}_{n-1} (y)
\eeq
where
\beq
\bar{\mS}_n = R_{2n+1} + R_{2n+2}
\eeq
and
\beq
\bar{\mR}_n = R_{2n+1} R_{2n}.
\eeq

Because of the extra factor of z associated with the odd series
the `barred' polynomials satisfy orthogonality condition

\beq
\int_0^{\infty} dy e^{-N^{\prime} [\mV_0 (y) -\bar{t}^{\prime} log y]}
\bar{\mP}_n (y) \bar{\mP}_m (y) = \bar{h}_n \delta_{n,m}
\eeq

where $\bar{t}^{\prime}=t+{1\over 2 N^{\prime}}$. This barred system
can be solved as a brick-wall problem. Original recurrence coefficients
are obtained by

\beq
R_{2n+1} = {1\over 2} \{ \mS_n + \sqrt{\mS^2_n-4\bar{\mR}_n} \}
\eeq

\beq
R_{2n} = {1\over 2} \{ \mS_n - \sqrt{\mS^2_n -4 \bar{\mR}_n} \}.
\eeq

Using $\mW_{n+1} + \mW_n + S_n \mY_n = {{2n+1+Nt}\over N}$ and
$S_n {\mW_{n+1}-\mW_n-{1\over N}}=-R_{n+1} \mY_{n+1}+R_n \mY_{n-1}$
(see ref. \cite{T92} for details and definitions of $\mW$ and $\mY$) 
we get

\beq
\mS_n = {{2n+1+t^{\prime}N^{\prime}}\over {\mu N^{\prime}}}
={{4n+1+tN}\over {\mu N}}
\eeq

\beq
\bar{\mS}_n = {{2n+1+\bar{t}^{\prime} N^{\prime}}
\over {\mu N^{\prime}}} = {{4n+3+tN}\over {\mu N}}
\eeq

\beq
\mR_n = {n(n+t^{\prime}N^{\prime})\over {\mu^2 N^{\prime 2}}}
={{2n(2n-1+tN)}\over {\mu^2 N^2}}
\eeq

\beq
\bar{\mR}_n = {{n(n+\bar{t}^{\prime} N^{\prime})}\over
{\mu^2 N^{\prime 2}}} = {{2n(2n+1+tN)} \over {\mu^2 N^2}}.
\eeq

(1). For the even set the orthogonality relation

\beq
\int_0^{\infty} dy e^{-N^{\prime}[\mV_0 (y) - t^{\prime} log y]}
\mP_n (y) \mP_m (y) = h_n \delta_{n,m} \nonumber \\
\eeq

simplifies to

\beq
\int_0^{\infty} dy e^{-N^{\prime} \mV_0 (y)} y^{N^{\prime} t^{\prime}}
\mP_n (y) \mP_m (y) = h_n \delta_{n,m}.\nonumber \\
\eeq

Now as

\beq
\mR_n = {{n(n+t^{\prime}N^{\prime})}\over {\mu^2 N^{\prime 2}}}
={{2n(2n-1+tN)}\over {\mu^2 N^2}}, \nonumber \\
\eeq

\beq
\mR_1 = {{2(2-1+tN)}\over {\mu^2 N^2}}
= {{2(1+tN)}\over {\mu^2 N^2}},\nonumber \\
\eeq

\beq
\mR_2 = {{4(4-1+tN)} \over {\mu^2 N^2}}\nonumber \\
\eeq
etc. Note the following 
\beq
\mR_n={h_n\over h_{n-1}}.
\eeq


Thus $h_n$ is
\beql
h_n &=& \mR_n h_{n-1}=\mR_n \mR_{n-1}h_{n-2} \nonumber \\
&=&\mR_n \mR_{n-1}\mR_{n-2}.........\mR_1 h_0 \nonumber \\
&=& {{n(n+t^{\prime}N^{\prime})}\over {\mu^2 N^{\prime 2}}}
{{(n-1)(n-1+t^{\prime}N^{\prime})}\over {\mu^2 N^{\prime 2}}}\nonumber \\
&& ...............\nonumber \\
&&{{(1+t^{\prime}N^{\prime})}
\over {\mu^2 N^{\prime 2}}}h_0 \nonumber \\
&=& {n!\over {(\mu N^{\prime})^{2n}}}
{{\Gamma(n+t^{\prime}N^{\prime}+1)}\over
{\Gamma(t^{\prime}N^{\prime}+1)}} h_0.
\eeql

Note:

\beq
\mP_n (y) = P_{2n} (z)
\eeq

for $n=0$

\beq
\mP_n (y) = P_0 (z) = 1.
\eeq

Then

\beql
h_0 &=& \int_0^{\infty} dy e^{-N^{\prime} \mV_0 (y)}
y^{N^{\prime} t^{\prime}}
\mP_0 (y) \mP_0 (y)\nonumber \\
&=& \int_0^{\infty} dy e^{-N^{\prime} 
\mV_0 (y)} y^{N^{\prime} t^{\prime}}.
\eeql

For 

\beq
\mV_0 (y) = \mu y,
\eeq

\beq
h_0=\int_0^{\infty} dy e^{-N^{\prime}\mu y}
y^{N^{\prime}t^{\prime}}.
\eeq

Defining

\beq
u=N^{\prime}\mu y,
\eeq

\beq
du=N^{\prime}\mu dy \nonumber
\eeq

we get 

\beql
h_0 &=& \int_0^{\infty} {du\over {N^{\prime}\mu}} e^{-u}
({u\over {N^{\prime}\mu}})^{N^{\prime} t^{\prime}}\nonumber \\
&=& {1 \over {(N^{\prime}\mu)^{N^{\prime}t^{\prime}+1}}} 
\Gamma (N^{\prime}t^{\prime}+1)
\eeql

and 

\beql
h_n &=& {{n! \Gamma(n+1+t^{\prime}N^{\prime})}\over
{(\mu N^{\prime})^{2n+N^{\prime}t^{\prime}+1}}}.\\
\eeql

Rewritting                   

\beq
\mP_n (y) = Q_n (u) 
\eeq

the orthogonality condition is

\beql
\int_0^{\infty} du u^{N^{\prime} t^{\prime}} e^{-u}
Q_n (u) Q_m (u) &=& (N^{\prime} \mu)^{N^{\prime}
t^{\prime}+1} h_n \delta_{n,m} \nonumber\\
&=& {{(n!)^2 \Gamma (n+1+t^{\prime} N^{\prime})}
\over {(\mu N^{\prime})^{2n} \Gamma (n+1)}} \delta_{n,m}.\nonumber
\eeql

Redefine $Q^{\prime}_n (u) = {{(\mu N^{\prime})^n}\over n!} Q_n (u)$.
Then

\beq
\int_0^{\infty} du u^{N^{\prime} t^{\prime}} e^{-u} Q_n^{\prime} (u)
Q_n^{\prime} (u) = {\Gamma (n+1+t^{\prime}N^{\prime})\over
\Gamma (n+1)} \delta_{n,m},
\eeq

comparing with the Associated Laguerre Polynomials $L^{(\alpha)}_n(x)$ 
we get

\beq
Q_n^{\prime} (u) = (-1)^n L_n^{N^{\prime}t^{\prime}} (u),
\eeq

then ($\alpha=N^{\prime}t^{\prime}$ Associated Laguerre Polynomials)

\beq
\mP_n (y) = Q_n (u) = {n!\over {(\mu N^{\prime})^n}}
(-1)^n L_n^{N^{\prime}t^{\prime}} (u)
\eeq

and 

\beql
\hat{\mP}_n (y) &=& {\mP_n (y) \over {\sqrt{h_n}}} \nonumber\\
&=& \sqrt{(\mu N^{\prime})^{N^{\prime}t^{\prime}+1}
\over {(n!)\Gamma(n+1+t^{\prime}N^{\prime})}}
(n!)(-1)^n L_n^{N^{\prime}t^{\prime}} (N^{\prime}\mu y).
\eeql

Thus the normalized even set of orthogonal polynomials are

\beql
\psi_{2n} (y) &=&
e^{-{N^{\prime}\over 2}[\mu y - t^{\prime}log y]}
\hat{\mP}_n (y)\nonumber \\
&=&
[{n!(N^{\prime}\mu)^{N^{\prime}t^{\prime}+1}
\over {\Gamma (n+1+N^{\prime}t^{\prime})}}]^{1\over 2}
y^{N^{\prime}t^{\prime}\over 2}
e^{-{N^{\prime}\mu y}\over 2} L_n^{N^{\prime}t^{\prime}}
(N^{\prime}\mu y).
\label{peven}
\eeql

(2). For odd set

\beq
\int_0^{\infty} dy e^{-N^{\prime}{[\mu y-\bar{t}^{\prime}log y]}}
\bar{\mP}_n (y) \bar{\mP}_m (y) = \bar{h}_n \delta_{nm}\nonumber
\eeq

where $\bar{t}^{\prime}=t+{1\over {2N^{\prime}}}$ and $N^{\prime}
\bar{t}^{\prime}={{Nt+1}\over 2}$. The orthogonality condition are

\beq
\int_0^{\infty} dy e^{-N^{\prime} \mu y}
y^{N^{\prime} \bar{t}^\prime} \bar{\mP}_n (y) \bar{\mP}_m (y)=
\bar{h}_n \delta_{nm}. \nonumber
\eeq

Note that

\beq
\bar{\mR}_n = {\bar{h}_n\over {\bar{h}_{n-1}}}.
\eeq

For

\beq
\bar{\mR}_n = {n(n+\bar{t}^{\prime})
\over {\mu^2 N^{\prime 2}}}.
\eeq

we get

\beql
\bar{h}_n &=& \bar{\mR}_n \bar{h}_{n-1}\nonumber\\
&=& \bar{\mR}_n \bar{\mR}_{n-1} \bar{\mR}_{n-2}..........
\bar{\mR}_1 \bar{h}_0\nonumber\\
&=& {{n!(n+\bar{t}^{\prime}N^{\prime})!}\over
{(\mu N^{\prime})^{2n} (\bar{t}^{\prime} N^{\prime})!}}
\bar{h}_0.
\eeql

Note:

$
\bar{\mP}_n (y) = z^{-1} P_{2n+1} (z)
$
and
$
\bar{\mP}_0 (y) = z^{-1} P_1 (z).
$

But $P_1 (z) = z$ hence $\bar{\mP}_0 (y) = z^{-1} z =1$. Therefore

\beql
\bar{h}_0 &=& \int_0^{\infty} dy e^{-N^{\prime} \mu y}
y^{N^{\prime}\bar{t}^{\prime}} \bar{\mP}_0 (y) \bar{\mP}_0 (y)\nonumber\\
&=& \int_0^{\infty} dy e^{-N^{\prime}\mu y}
y^{N^{\prime}\bar{t}^{\prime}}.
\eeql

Defining

\beq
u = N^{\prime} \mu y 
\eeq

and 

\beq
du = N^{\prime} \mu dy, \nonumber
\eeq

we get 

\beql
\bar{h}_0 &=& \int_0^{\infty} {du\over {N^{\prime}\mu}}
e^{-u} ({u\over {N^{\prime} \mu}})^{N^{\prime}\bar{t}^{\prime}}\nonumber\\
&=& {1\over {(N^{\prime}\mu)^{N^{\prime}\bar{t}^{\prime}+1}}}
\int_0^{\infty} du e^{-u} u^{N^{\prime} \bar{t}^{\prime}}.
\eeql

So that

\beq
\bar{h_n} = {{\Gamma (n+1) \Gamma (n+1+\bar{t}^{\prime} N^{\prime})}
\over {(N^{\prime} \mu)^{2n+N^{\prime}\bar{t}^{\prime}+1}}}.
\eeq

Now let $\bar{Q}_n (u) = \bar{\mP}_n (y)$. Therefore

\beql
\int_0^{\infty} du e^{-u} u^{N^{\prime}\bar{t}^{\prime}}
\bar{Q}_n (u) \bar{Q}_m (u) &=&
(N^{\prime}\mu)^{N^{\prime}\bar{t}^{\prime}+1} \bar{h}_n \delta_{nm}
\nonumber\\
&=& {{\Gamma(n+1) \Gamma(n+1+\bar{t}^{\prime}N^{\prime})}\over
{(\mu N^{\prime})^{2n}}} \delta_{nm}\nonumber\\
&=& ({{n!}\over {(\mu N^{\prime})^n}})^2
{{\Gamma (n+1+\bar{t}^{\prime}N^{\prime})}\over {n!}} \delta_{nm}.
\eeql

Define $\bar{Q}_n^{\prime} (u)= {(\mu N^{\prime})^n\over n!}
\bar{Q}_n (u)$ then

\beq
\int_0^{\infty} du e^{-u} u^{N^{\prime}\bar{t}^{\prime}}
\bar{Q}^{\prime}_n (u) \bar{Q}_m^{\prime} (u) =
{\Gamma (n+1+\bar{t}^{\prime} N^{\prime}) \over {n!}} \delta_{nm}.
\eeq

For

\beql
\bar{Q}_n^{\prime} (u) &=& {(\mu N^{\prime})^n\over n!} Q_n (u)\nonumber\\
& = & (-1)^n L_n^{N^{\prime}\bar{t}^{\prime}} (u) \nonumber\\
\eeql

where $\alpha=N^{\prime}\bar{t}^{\prime}$. Now

\beq
\bar \mP_n (y) = \bar{Q}_n (u) = {n! (-1)^n \over (\mu N^{\prime})^n}
L^{N^{\prime}\bar{t}^{\prime}}_n (N^{\prime}\mu y).
\eeq

hence

\beql
\hat{\bar{\mP}}_n (y) & = & {\bar{\mP}_n (y)\over 
{\sqrt{\bar{h_n}}}}\nonumber\\
&=& \sqrt{(\mu N^{\prime})^{N^{\prime}\bar{t}^{\prime}+1} n!
\over {\Gamma(n+1+\bar{t}^{\prime}N^{\prime})}}
(-1)^n L_n^{N^{\prime}\bar{t}^{\prime}} (N^{\prime}\mu y).
\eeql

Thus the normalized odd orthogonal polynomials are

\beql
\psi_{2n+1} (y) &=&
e^{-{N^{\prime}\over 2}[\mu y - \bar{t}^{\prime}log y]}
\hat{\bar{\mP}}_n (y)\nonumber \\
&=&
[{n!(N^{\prime}\mu)^{N^{\prime}\bar{t}^{\prime}+1}
\over {\Gamma (n+1+N^{\prime}\bar{t}^{\prime})}}]^{1\over 2}
y^{N^{\prime}\bar{t}^{\prime}\over 2}
e^{-{N^{\prime}\mu y}\over 2} L_n^{N^{\prime}\bar{t}^{\prime}}
(N^{\prime}\mu y).
\label{podd}
\eeql

The important lesson to learn from this exercise is that 
$\alpha={{(Nt-(-1)^{n^\prime})}\over 2}$ for $n^{\prime}$ an
integer which is even or odd and $x=N^{\prime}\mu y$
for the orthogonal polynomials, which are proportional to the
generalized Laguerre polynomials $L^{(\alpha)}_n (x)$,
of the Gaussian Penner Matrix Model.

\sect{The Old Asymtotic Formula for Associative Laguerre Ensemble}

Consider the asymptotic formula given in Szego's book on
orthogonal polynomials. The formulas of Plancherel-Rotach type for
Laguerre polynomials for $\alpha$ arbitrary and real, $\epsilon$
fixed positive number for $x=(4n+2\alpha+2)\cos^2\phi$,
$\epsilon \le \phi \le {\pi\over 2} - \epsilon n^{-1\over 2}$ 
are given below

\beql
&&e^{-x\over 2} L_n^{(\alpha)} (x) =  (-1)^n (\pi sin \phi)^{-1\over 2}
x^{-{\alpha\over 2}- {1\over 4}} n^{{\alpha\over 2}-{1\over 4}}\nonumber\\
&&\{ Cos [ (n+{{(\alpha+1)}\over 2})(sin 2\phi - 2 \phi) + {\pi\over 4}]
+(nx)^{-1\over 2} O(1) \}.
\eeql

For the Gaussian Penner model with $\alpha={{(Nt-(-1)^{n^\prime})}\over 2}$ 
and $x=N^{\prime}\mu y$

\beql
&&e^{-x\over 2} L_n^{(\alpha)} (x) = (-1)^n (\pi sin \phi)^{-1\over 2}
x^{{-\alpha\over 2}-{1\over 4}} n^{{\alpha\over 2}-{1\over 4}}\nonumber\\
&&\{ Cos [ (n+{1\over 2})(sin 2\phi - 2\phi) +
{{(Nt-(-1)^{n^\prime})}\over 4} (sin 2\phi - 2\phi) + {\pi\over 4}]\nonumber\\
&& + (nx)^{-1\over 2} O(1) \}.
\eeql

The combined asymptotic ansatz for the even and odd results 
eq. (\ref{peven}) and eq. (\ref{podd}) for the Gaussian Penner 
model in this asymtotic region is ($n^\prime$ stands for both 
even $2n$ and odd $2n+1$ integers)

\beql
&&\psi_{n^\prime} (x) =  e^{-x\over 2} L_n^{(\alpha)} (x)
= (-1)^n (\pi sin \phi)^{-1\over 2} x^{{-\alpha\over 2} - {1\over 4}}
n^{{\alpha\over 2}-{1\over 4}} \nonumber \\
&&\{ Cos [ (n+{1\over 2}+{Nt\over 4})
(sin 2\phi - 2\phi) + {\pi\over 4} - {(-1)^{n^\prime}\over 4}
(sin 2\phi -2\phi) ] \nonumber\\
&&+ (nx)^{-{1\over 2}} O(1) \}
\eeql

Note that the term ${(-1)^{n^\prime}\over 4}(sin 2\phi -2\phi)$ is
not the extra term (with $\eta$ see ref. \cite{D97,BD99} or  
eq. (\ref{fzpec}) below) that was 
found for the asymptotic ansatz in the double well problem.

\sect{The New Asymptotic Formula for Associative Laguerre Ensemble}
\label{alag}

Let us start with an expression which has a well defined $\alpha=0$ limit. 
Using

\beq
{d\over dx} f(x) = {d\over du} f(u+x)|_{u=0}
\eeq

we can write using Cauchy's theorem

\beql
L^{(\alpha)}_n (x) &=& {e^x\over n!} x^{-\alpha}     
({d\over du})^n [(x+u)^{n+\alpha} e^{-(x+u)}]_{u=0}\nonumber\\
&=& {x^{-\alpha}\over n!} ({d\over du})^n
[(x+u)^{(n+\alpha)} e^{-(u)}]|_{u=0}\nonumber\\
&=& x^{-\alpha} \int_c {dz\over 2i\pi} {1\over z^{n+1}}
(z+x)^{n+\alpha} e^{-z}\nonumber\\
\eeql

in which the countour C is a small circle around the origin
($|z|<x$) change $z\rightarrow nz$ ($|z|<{x\over n}$)

\beql
L^{(\alpha)}_n (x) &=& ({x\over n})^{-\alpha} \int_c {dz\over 2i\pi}
{1\over z^{n+1}} (z+{x\over n})^{n+\alpha} e^{-nz}\nonumber \\
&=& ({x\over n})^{-\alpha} \int_c {dz\over 2i\pi} {1\over z}
e^{-n f(z)}
\eeql

\beq
f(z) = z + log z - (1+{\alpha\over n}) log (z+{x\over n})
\eeq

we explore ${\alpha\over n}$ and ${x\over n}$ finite.
In this representation the limit $\alpha=0$ is well defined.
C is a circle around the origin. The saddle point is given
by an expression

\beq
z^2 + z ({{x-\alpha}\over n}) + {x\over n} = 0.
\eeq

With the parametrization

\beq
{\alpha-x\over n} = 2 \sqrt{x\over n} cos\phi
\eeq

\beq
z_0 = \sqrt{x\over n} e^{i\phi}
\eeq

and $\bar{z_0}$ are the saddle points. This is valid in the range
$|{\alpha-x\over n}| < 2 \sqrt{x\over n}$; if the parameters are
such that this inequality is reversed, there is only one real
saddle-point. This is where the new saddle points will have to be 
taken into account. 

If there is a saddle-point in the integral

\beq
I=\int_c {dz\over 2i\pi} {1\over z} e^{-n f(z)}
\eeq

at $z_0$ then

\beq
I \approx {e^{-n f(z_0)} \over z_0} \sqrt{2\pi\over n |f''(z_0)|}
e^{-i\theta\over 2}
\eeq

with $f''(z_0)=|f''(z_0)|e^{i\theta}$. Indeed expand $f(z)$ around
$z_0$

\beq
f(z) = f(z_0) + {1\over 2} (z-z_0)^2 |f''(z_0)| e^{i\theta}
\eeq

path is $ z-z_0 = x e^{{-i\theta}\over 2}$

\beq
I \approx e^{-i{\theta\over 2} - n f(z_0)} \int_{-\infty}^{+\infty}
dx  e^{-{n\over 2} x^2 |f''(z_0)|}.
\eeq

Here we need to add the contributions of $z_0$, $\bar{z}_0$ and
the other saddle points. Therefore if $f(z_0)=A+iB$

\beq
I \approx e^{-nA} \sqrt{2\pi\over x |f''(z_0)|}
2 \cos (nB+\phi+{\theta\over 2})(x)
\eeq

with

\beq
f''(z_0)= \sqrt{n\over x} {2i ~~ sin \phi ~~ e^{-i\phi}\over
\sqrt{x\over n}+e^{i\phi}}
\eeq

and $\theta$ is it's phase. The $(-1)^n$ would only show up in B.
This is a new asymptotic expansion of the generalized Laguerre
ensemble in a novel asymptotic regime. It would be nice to
have a physical picture of this asymptotic regime.

On simplifying the above expression it will give the 
asymtotic ansatz found in ref.\cite{D97} for the double well
matrix model. Following ref. \cite{D97} for $N$ large but 
$N-n\approx O(1)$ and $x$ lying in the two cuts the asymptotic ansatz
for the orthogonal polynomials of the Gaussain Penner Matrix Model
can be approximated by 

\beq
\psi_n (x) = {1\over \sqrt{f}} 
\left [ \cos ( N \zeta - (N-n)\phi + \chi + (-1)^n \eta )(x) 
+ O ({1\over N}) \right ]
\eeq

where $f, \zeta, \phi, \chi$ and $\eta$ are functions of $x$ and
are given by

\beql
f(x) &=& {\pi\over 2 x} {(b^2-a^2)\over 2} 
\sin 2\phi (x) \nonumber \\
\zeta^{\prime} (x) &=& - \pi \rho (x) \nonumber \\
\cos 2 \phi (x) &=& {x^2-{(a^2+b^2)\over 2}\over 
{(b^2-a^2)\over 2}} \nonumber \\
\cos 2 \eta (x) &=& b {\cos \phi (x) \over x} 
\nonumber \\
\sin 2 \eta (x) &=& a {\sin \phi (x) \over x}
\nonumber \\
\chi (x) &=& {1\over 2} \phi (x) - {\pi\over 4}
\label{fzpec}
\eeql
with $a^2$ and $b^2$ as given in sec. (2) for the 
Gaussian Penner model.

All the corresponding correlation functions of 
this model will be as obained in ref. \cite{D97} and ref. \cite{BD99}. 
Following ref. \cite{BD99} and using the contour of integration
for the Gaussian Penner model, see Fig. \ref{pennerc}, the 
the smoothed density-density correlation function can be derived  
in the thermodynamic limit and is an oscillating function of $N$: 

\begin{figure}     
\leavevmode
\epsfxsize=4in
\epsffile{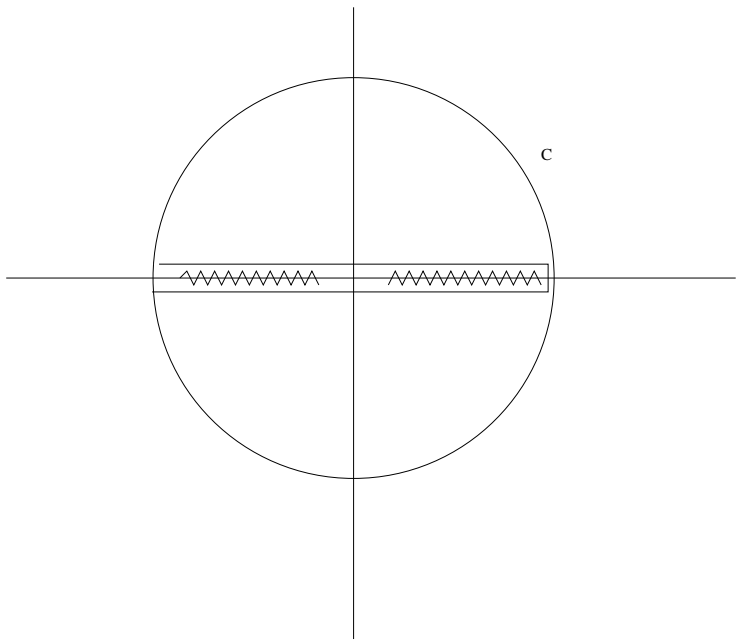}
\caption{The contour of integration for the two-cut Gaussian 
Penner model.}
\label{pennerc} 
\end{figure}

\beql
2 \pi^2 N^2 \rho^c_2 (\lambda,\mu) = 
{{\epsilon_{\lambda}\epsilon_{\mu}} \over {\beta \sqrt{|\sigma(\lambda)|}
\sqrt{|\sigma(\mu)|}}} {1\over {(\mu -\lambda)^2}}\nonumber \\
\left( \lambda \mu (\lambda \mu - a^2 - b^2) + a^2 b^2
+ (-1)^N ab (\mu-\lambda)^2 \right).\nonumber \\
\label{dd}
\eeql

Here  for the symmetric Gaussian Penner Model
$\sigma(z)=(z^2-a^2)(z^2-b^2)$, 
$a^2={(2+t)\over \mu}+{2\over \mu}\sqrt{(1+t)}$, $b^2={(2+t)\over \mu}
-{2\over \mu}\sqrt{(1+t)}$,
$\epsilon_{\lambda}=+1$ for $b<\lambda<a$, $\epsilon_{\lambda}=-1$ for
$-a<\lambda<-b$ and $\beta=1,2,4$ depending on whether $M$ the matrix
is real orthogonal, hermitian or self-dual quartonian. 

Using this expression in the 
formulas for the mesoscopic fluctuation \cite{B94} 
would give rise to terms depending on $N$. 
This may be observed in single electron experiments on mesoscopic
samples which have gaps in their eigenvalue spectrum. Work to explicitly
obtain all these expressions in the Chiral Matrix Model in this asymptotic
regime is in progress.

\sect{Conclusion}

Finally let us note that in matrix models
when the number of connected components
for the support of the eigenvalues
changes, one finds a new universality class for the correlators
which has been extended here to include the non-polynomial potentials.
It is not completely obvious that
it is legitimate to use the simple one cut function
in the application to mesoscopic fluctuations as the 
correlation function are different for these gapped random
matrix models because of single electron tunneling effects.
Further questions of symmetry breaking in these singular 
random matrix models are open questions to be explored 
in the future.

\vskip 5mm

\setcounter{equation}{0}
\renewcommand{\theequation}{A.\arabic{equation}}
\noindent

\vskip 5mm

\begin{flushleft}
\underline{Acknowledgements}:\hfill\break

ND thanks Prof. E. Br\'ezin for discussions. ND also thanks
the Abdus Salam International Centre for Theoretical Studies
for hospitality and facilities.

\end{flushleft}

\vskip 5mm


\newcommand{\NP}[3]{{\it Nucl. Phys. }{\bf B#1} (#2) #3}
\newcommand{\PL}[3]{{\it Phys. Lett. }{\bf B#1} (#2) #3}
\newcommand{\PR}[3]{{\it Phys. Rev. }{\bf #1} (#2) #3}
\newcommand{\PRL}[3]{{\it Phys. Rev. Lett. }{\bf #1} (#2) #3}
\newcommand{\IMP}[3]{{\it Int. J. Mod. Phys }{\bf #1} (#2) #3}
\newcommand{\MPL}[3]{{\it Mod. Phys. Lett. }{\bf #1} (#2) #3}
\newcommand{\JP}[3]{{\it J. Phys. }{\bf A#1} (#2) #3}

\end{document}